\documentclass[final,5p,times,twocolumn]{elsarticle}

\usepackage{amssymb}

\usepackage[utf8]{inputenc}
\usepackage{multirow}
\usepackage[bottom]{footmisc}
\usepackage{setspace}
\usepackage{amssymb}
\usepackage{amsmath}
\usepackage{amsthm}
\usepackage{color}
\usepackage{graphicx}
\usepackage{appendix}
\usepackage{epstopdf}
\usepackage{float}
\usepackage{url}
\usepackage{subfig}
\usepackage{multirow}
\usepackage{graphicx}
\usepackage{booktabs}
\long\def\symbolfootnote[#1]#2{\begingroup%
\def\thefootnote{\fnsymbol{footnote}}\footnote[#1]{#2}\endgroup}
\usepackage{soul}
\sethlcolor{yellow}
\newtheorem{theorem}{Theorem}[section]
\newtheorem{cor}[theorem]{Corollary}
\newtheorem{lem}[theorem]{Lemma}

\allowdisplaybreaks[4]
\usepackage{amsthm,amsmath,amssymb,amscd}

\usepackage{algorithm}
\usepackage[noend]{algpseudocode}
\usepackage{mleftright}

\usepackage{array}
\usepackage{url}
\journal{Journal of Biomedical Informatics}
\biboptions{sort&compress}
\usepackage{lipsum}
\begin{document}
\setlength{\abovedisplayskip}{2pt}
\setlength{\belowdisplayskip}{2pt}
\mleftright
\begin{frontmatter}
 \author[r]{Aven Samareh\corref{cor1}}
\ead{asamareh@uw.com}
\author[r]{Shuai Huang}
\ead{shuaih@uw.edu}
\address[r]{Industrial \& Systems Engineering Department, University of Washington, Seattle, WA, 98195, USA}
\cortext[cor1]{Corresponding authors.}
\title{UQ-CHI: An Uncertainty Quantification-Based Contemporaneous Health Index for Degenerative Disease Monitoring}

\begin{abstract}
Developing a knowledge-driven contemporaneous health index (CHI) that can precisely reflect the underlying patient condition across the course of the condition’s progression holds a unique value, like facilitating a range of clinical decision-making opportunities. This is particularly important for monitoring degenerative conditions such as Alzheimer's disease (AD), where the condition of the patient will decay over time. Detecting early symptoms and progression signs, and continuous severity evaluation, are all essential for disease management. While a few methods have been developed in the literature, uncertainty quantification of those health index models has been largely neglected. To ensure the continuity of the care, we should be more explicit about the level of confidence in model outputs. Ideally, decision-makers should be provided with recommendations that are robust in the face of substantial uncertainty about future outcomes. In this paper, we aim at filling this gap by developing an uncertainty quantification based contemporaneous longitudinal index, named UQ-CHI, with a particular focus on continuous patient monitoring of degenerative conditions. Our method is to combine convex optimization and Bayesian learning using the maximum entropy learning (MEL) framework, integrating uncertainty on labels as well. Our methodology also provides closed-form solutions in some important decision making tasks, e.g., such as predicting the label of a new sample. Numerical studies demonstrate the effectiveness of the proposed UQ-CHI method in prediction accuracy, monitoring efficacy, and unique advantages if uncertainty quantification is enabled in practice.
\end{abstract}

\end{frontmatter}

\section{Introduction} \label{introduction}
The effective monitoring of degenerative patient conditions represents a significant challenge in many clinical decision-making problems and has given rise to the development of numerous mathematical and computational models \cite{brownell1999dopamine,gratwicke2017early,llano2017multivariate,chen2014credit}. Developing a knowledge-driven contemporaneous health index (CHI) that can precisely reflect the underlying patient condition across the course of the condition’s progression holds a unique value, like facilitating a range of clinical decision-making opportunities \cite{spring2013healthy,rivera2012optimized,deshpande2014control}, enhancing the continuity of care, and facilitating communications between clinicians, healthcare providers, and patients. It will also be a crucial enabling factor for the development of many envisioned AI systems to implement adaptive interventions for better healthcare management, given a representation of the dynamic evolution of the patient's condition. 

Thus, to ensure continuity of care, we should be more explicit about our level of confidence in model outputs. Ideally, decision-makers should be provided with recommendations that are robust in the face of substantial uncertainty about future outcomes. However, computational models are an abstraction of clinical observations, as such, they are usually built on analytically tractable assumptions that may simplify the real-world problem. Also, most of these models are estimated from imperfect data, subjecting them to all kinds of statistical errors. An approach that yields only a single prediction doesn't adequately reflect any uncertainty, neither in the empirical data nor the estimated parameters \cite{allmaras2013estimating}. As a result, the outcomes from such mathematical models may not be consistent with the clinical observations. Uncertainty is an unavoidable feature that affects prediction capabilities in real-world domains such as healthcare \cite{hoffman1994propagation, meghdadi2017brain}, manufacturing \cite{montomoli2015uncertainty, nannapaneni2014uncertainty}, signal processing \cite{reynders2016uncertainty, nobari2015uncertainty}, and etc. A certain amount of uncertainty is always involved in decision-making systems that do not encounter samples when the experimental data are insufficient to calibrate. In such cases, there is always a chance that the model parameters be determined unambiguously even in the existence of complex mathematical models. In clinical predictions, it is necessary to deal with such uncertainty in an effective manner, because if the model parameters are not well constrained, the resulting predictions may represent an unacceptable degree of posterior uncertainty. What is more, while most existing models in patient monitoring generate one single prediction without telling confidence level, uncertainty quantification could tell us on which samples we may not be ready to act based on the model. Therefore, to develop a reliable model for a clinically relevant prediction, uncertainty quantification is a much-needed capacity \cite{collis2017bayesian,biglino2017computational,bozzi2017uncertainty}. 

A number of patient monitoring index approaches have been developed in the literature. A standard formulation of these health indices is to use weighted sum models (e.g., regression models), and combine multiple static clinical measurements to predict the disease condition. For example, there exist many risk score models to predict AD by using multi-modality data integration methods \cite{liu2013data,yuan2012multi,zhang2011multimodal} to combine neuroimaging data \cite{weiner2013alzheimer,weiner20152014}, genomics data \cite{biffi2010genetic}, clinical data \cite{reitz2010summary}, etc. There are a few approaches that have formulated the decline of AD-related score over time as a multi-task learning model \cite{zhou2013modeling,zhou2012modeling}. These existing efforts have been limited to combining static data rather than longitudinal data. Besides, these data are usually sampled at irregular time points, which adds in another layer of complexity to the modeling efforts. Our problem's objective is fundamentally different from the existing risk score models; we focus on developing the contemporaneous health index (CHI) that can fuse irregular multivariate longitudinal time series dataset to quantify the severity of degenerative disease conditions that are required to fit the monotonic degradation process of the disease condition. For example, in our previous work \cite{ samareh2018dl} to address the patient heterogeneity, we developed a dictionary learning-based contemporaneous health index for degenerative disease monitoring, called DL-CHI, that leveraged the knowledge of the monotonic disease progression process to fuse the data by integrating CHI with dictionary learning. The basic idea of DL-CHI was learning individual models via the CHI formulation, and then rebuilding the model parameters of each patient's models through a supervised dictionary learning. However, both CHI and DL-CHI frameworks only generate one single prediction value for a sample and ignore the sampling uncertainty (i.e., it is common in healthcare that the label information is usually obtained by subjective methods which are subject to uncertainty). Therefore, if we could enable CHI to conduct uncertainty quantification and incorporate the uncertainty in labels in its modeling, we can widen its applicability in real-world contexts. The main objective of this paper is to develop a framework that can focus on the contemporaneous health index (CHI) developed in \cite{huang2017chi}, and can further equip CHI with uncertainty quantification capacity.

In this paper, we develop the uncertainty quantification based contemporaneous longitudinal index, named UQ-CHI, with a particular focus on continuous patient monitoring of degenerative conditions. Our method is to combine convex optimization and Bayesian learning using the maximum entropy learning (MEL) framework, integrating uncertainty on labels as well. The basic idea of MEL is to identify the distribution of the parameters of a statistical model that bears the maximum uncertainty, a principle that is conservative and robust \cite{mackay2003information,izenman2008modern,phillips2006maximum}. It has been investigated in a few machine learning models \cite{jaakkola2000maximum,sun2013multi,chao2019semi,zhu2018semi} as well. For example, in \cite{jaakkola2000maximum}, MEL was used to learn a distribution of the parameters in the support vector machine model rather than a single vector of the parameters. This distribution of the parameters could help us evaluate the uncertainty of the learned support vector machine model and translate into the uncertainty of predictions.

\begin{figure*}[!ht]
\includegraphics[scale=0.1]{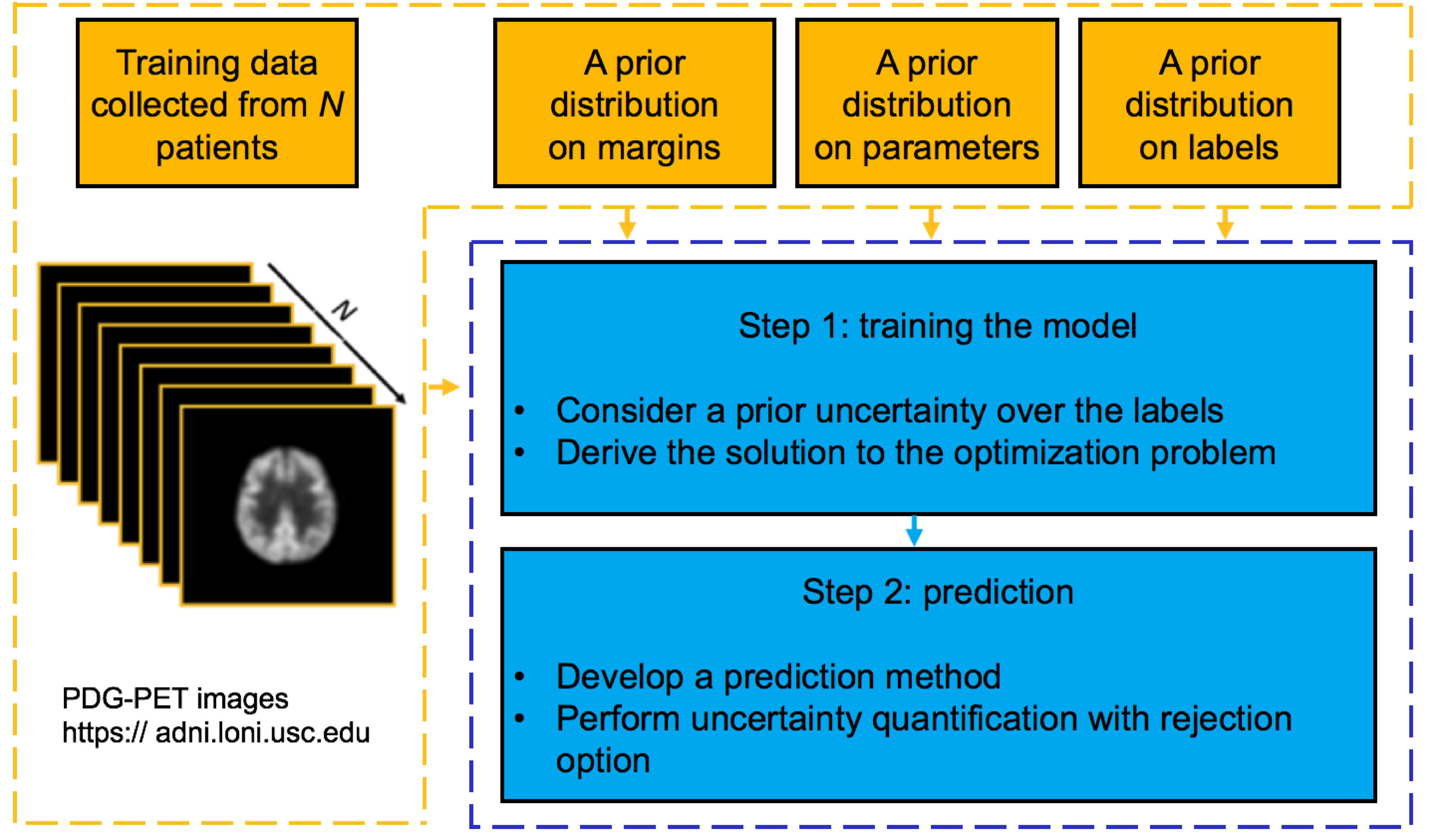}
\centering
\caption{A conceptual overview of the UQ-CHI method}
\label{figure_1}
\end{figure*}

To adapt the MEL formulation and to develop UQ-CHI, few challenges should be addressed. The objective function of MEL, as its distinct feature, bears the full spirit of maximum entropy: no matter what is the model, we are studying, the learning objective of MEL is to learn the distribution model of the parameters of the model that has the maximum entropy. If there is a prior distribution of the parameters, the Kullback–Leibler divergence could be used to extend this idea. In our case, the design of the prior distribution should be studied to account for label uncertainties. Besides the objective function, the MEL encodes information from the data into constraints, e.g., if the model is for classification, for each sample, there would be a constraint that the expectation of the prediction over the distribution of the parameters should match the observed outcome on this sample. In our case, we will derive the constraints from the CHI model and integrate with the MEL framework. In detail, we consider two steps in our method, i.e., training and prediction. In the training step, we consider a prior uncertainty over the labels to handle uncertain or incomplete labels. Then we derive a solution to the optimization problem by using a specific prior formulation. In the second step, we develop a prediction method, with a rejection option method, for new samples with the obtained uncertainty quantification capacity. A distinct feature of our model is that it provides a closed-form solution for predicting the label of a new example. The whole pipeline of this UQ-CHI model is shown in Figure \ref{figure_1}. 

The remainder of this paper is organized as follows: in Section \ref{related works}, we will review related literature in modeling the contemporaneous health index for degenerative conditions and the MEL framework. In Section \ref{proposed work}, the UQ-CHI framework will be presented. In Section \ref{numerical studies}, we will implement and evaluate the UQ-CHI using a simulated dataset. We then continue the numerical analysis with a real-world application on Alzheimer's disease dataset in Section \ref{ADNI}. We will conclude the study in Section \ref{concl}. Note that, in this paper, we use lowercase letters, e.g., x, to represent scalars, boldface lowercase letters, e.g., \textbf{v}, to represent vectors, and boldface uppercase letters, e.g., \textbf{W}, to represent matrices.
\section{Related works}\label{related works}
In this section, we will first briefly present the basic formulation of the contemporaneous health index (CHI) model, and its extension, the dictionary learning based contemporaneous health index (DL-CHI), then we will present the proposed model: the UQ-CHI. 
\subsection{The CHI model}\label{CHI}
The CHI model is developed in \cite{huang2017chi} which exploits the monotonic pattern of disease over the course of progression to improve further the data fusion of multivariate clinical measurements taken at irregular time points. The CHI framework was inspired by the common characteristics of degenerative conditions (e.g., AD) that often cause irreversible degradation. For example, in AD, to measure the degradation of the neural systems a number of biomarkers were developed, including neuroimaging modalities such as PET and MRI scans \cite{mueller2005alzheimer,petrella2003neuroimaging}. For example, MRI scans show a decline in the brain volume over time along with the disease progression. The same phenomenon could be observed on the PET scans when there is a persistent shrinkage of metabolic activities. Such monotonic patterns indicate that once the disease progression started, it tends to deteriorate over time increasingly. The task of CHI is to translate multivariate longitudinal and irregular clinical measurements into a contemporaneous health index $h_{n,t}$ to capture the patient’s condition changing over the course of progression. Note, clinical measurements for each patient could be taken with different length of time and at different time locations. Targeting degenerative conditions, CHI is designed to be monotonic, i.e., $h_{n,t_{1}}\geq h_{n,t_{2}}$ if $t_{1}\geq t_{2}$, while higher index represents a more severe condition. CHI is a latent structure; hence, clinical variables associated with it should be measured over time to facilitate data for learning the index.

Let, $ \mathbf{x}_{n, t} = \left[x_{n,1,t},\ldots, x_{n,d,t}\right]^ T\in \mathbb{R} ^{d}$, denote a training set of $N$ patients. Each measurement $x_{n,i,t}$, is the value of the $i$th variable for the $n$th subject in a given time $t$, where $t\in \left\{ 1,\ldots,T_{n}\right\}$ is the time index. our goal is, given a training set, convert each measurement $ \mathbf{x}_{n,t}$ into an health index $h_{n,t}$, which requires a mathematical model of $h_{n,t} = f( \mathbf{x}_{n,t})$. For simplicity, multivariable form of the hypothesis function $h_{n,t}$ was studies in \cite{huang2017chi}, i.e., $h_{n,t} = \mathbf{x}_{n,t}.\mathbf{w}$, where $\mathbf{w} \in R^{d}$ is a vector of weight coefficients that combines the $d$ variables. The total number of positive and negative samples is shown by $N^+$ and $N^-$ respectively, i.e., $N^+:=|\{n|y_n=1\}|$ and $N^-:=|\{n|y_n=-1\}|$. The formulation of the CHI learning framework is shown in below:
\begin{subequations}  \label{eq:original}
 \begin{align}
\min_{\mathbf{w},b}\quad & {\frac{1}{2}} \|\mathbf{w}\|^2 + \label{eq:original_a}\\ 
 &\beta\sum_{n\in \{1,\ldots,N\}} \max\bigg(0, 1-y_n(  \mathbf{x}_{n, T_n}^\top \mathbf{w}+b)\bigg) +\label{eq:original_b}\\
 &\alpha\sum_{\substack{n\in \{1,,\ldots,,N\} \\ t\in \{1,\ldots, T_n-1\}}} \max\bigg(0, 1-\mathbf{z}_{n, t}^\top \mathbf{w}\bigg) 
+ \label{eq:original_c}\\
&{\frac{\lambda}{2}} \Bigg({\frac{1}{N^+}} \sum_{n\in \{N^+|y_n=1\}} \bigg((  \mathbf{x}_{n, T_{n}}-\bar{  \mathbf{x}}^+_{T_{n}})^T \mathbf{w}\bigg)^2 \Bigg)+\label{eq:original_d}\\
&{\frac{\lambda}{2}} \Bigg({\frac{1}{N^-}} \sum_{n\in \{N^-|y_n=-1\}} \bigg((  \mathbf{x}_{n, T_{n}}-\bar{  \mathbf{x}}^-_{T_{n}})^T \mathbf{w}\bigg)^2 \Bigg) \label{eq:original_e} +\\
&\gamma \|\mathbf{w}\|_1.\label{eq:original_f}
 \end{align}
\end{subequations}

Items in \eqref{eq:original} can be explained as follows:
\begin{itemize}
\item The first term \eqref{eq:original_a} and the second term \eqref{eq:original_b} are derived from a general formulation of support vector machine (SVM). These two terms are used to enhance the discriminatory power of CHI by utilizing the label information. Here, $y_n \in \{1, -1\}$ is the label of the $n$th sample that indicates if the $n$th subject has the disease or not. 
\item To accommodate the monotonic pattern of disease progression, and to enforce the monotonicity of the learned health index, the term \eqref{eq:original_c} is invented, i.e., $h_{n,t_{1}}\geq h_{n,t_{2}}$ if $t_{1}\geq t_{2}$. Here, $\mathbf{z}_{n,t}$ is the difference of two successive data vectors $\mathbf{z}_{n,t}:= \mathbf{x}_{n,t+1}- \mathbf{x}_{n,t}$. 
\item To encourage the homogeneity of CHI within the group that has the same health status terms \eqref{eq:original_d} and \eqref{eq:original_e} are invented. Here, $\mathbf{\bar{x}}^+_{T_n}$ and $\mathbf{\bar{x}}^-_{T_n}$ represent the center of data vectors at time $T_n$ for all positive and negative samples, respectively, that are, 
\begin{align*}
\bar{ \mathbf{x}}^+_{T_n}:=&{\frac{1}{N^+}}\sum_{n\in \{n|y_n=1\}} \mathbf{x}_{n, T_n}
\\
\bar{ \mathbf{x}}^-_{T_n}:=&{\frac{1}{N^-}}\sum_{n\in \{n|y_n=-1\}} \mathbf{x}_{n, T_n}.
\end{align*}
\item To encourage sparsity of the features, $L_1$-norm penalty is used as shown in the last term \eqref{eq:original_f}.
\end{itemize}

The CHI formulation can be solved by using the block coordinate descent algorithm that is illustrated in \cite{huang2017chi}. Note, the CHI formulation generalizes many existing models, such as SVM, sparse SVM, LASSO, etc. 
\subsection{The DL-CHI model}\label{DL-CHI}
CHI formulation is designed for learning a model for the average of a population, and thus, ignores the patient heterogeneity. Patients who suffer from AD have very heterogeneous progression patterns \cite{cummings2000cognitive,folstein1989heterogeneity,friedland1988alzheimer}. Building a personalized model on an individual basis could be used to consider the heterogeneity. However, such models require a significant amount of labeled training samples, which is not feasible in such clinical settings. Towards this goal, the DL-CHI approach was further developed in \cite{samareh2018dl} by integrating CHI with dictionary learning \cite{olshausen1996emergence,cummings2000cognitive}.  Dictionary learning algorithms reconstruct the input signals as an approximated signal via a sparse linear combination of a few dictionary elements or basis \cite{wright2009robust} (each column of the dictionary represents a basis vector). Dictionary learning algorithms can reveal the hidden structures in the data (in a similar spirit as principal component analysis) by spanning the space of a personalized model and capturing patient heterogeneity. They play a role in the regularization of the model learning, in a way that each dictionary basis vector can be viewed as the numerical representations of patient heterogeneity. Thus, DL algorithms can improve the classification performance. Translating this wisdom into DL-CHI, the basic idea is first to learn individual models through the CHI formulation, and then, reconstruct the model parameters of the individual learned models via supervised dictionary learning. As such, each model is represented as a sparse linear combination of the basis vectors. Numerous experiments in both simulated and real-world data have shown the effectiveness of DL-CHI in creating personalized CHI models.

Despite accounting the patient heterogeneity, DL-CHI ignores the sampling uncertainty, therefore limits its applicability in real-world applications. Thus, this motivates us to enable CHI to conduct uncertainty quantification.
\subsection{The MEL formulation} \label{MED}
As mentioned in Section \ref{introduction}, MEL formulation has a distinct objective function that aims to learn the distribution of the parameters of a model that encodes maximum uncertainty (i.e., evaluated by the entropy concept). It also has constraints that encode information from the data, e.g., if the model is for classification, for each sample, there would be a constraint that the expectation of the prediction over the distribution of the parameters should match the observed outcome on this sample. To further illustrate some details, one typical application of the MEL is the maximum entropy discrimination (MED) method that focuses on the application of MEL on classification models.

Let’s consider a binary classification problem, where the response variable $y$ takes values from $\left\{ +1,-1\right\}$. Let $  \mathbf{x}_n =[  \mathbf{x}_1,\dots,  \mathbf{x}_n]$ be an input feature vector and $\mathcal{D}\left(   \mathbf{x}_{n}|\mathbf{w} \right)$ be a discriminant function parameterized by $\mathbf{w}$, and $\mathbf{\gamma}$ e.g., $\mathcal{D}(  \mathbf{x}_n|\mathbf{w}) = \mathbf{w}^T   \mathbf{x}_n $. The training set is defined by $D=\left\{   \mathbf{x}_{n},y_{n}\right\} ^{N}_{n=1}$ and the hinge loss is defined as $h(x) = \max \left( 0,y_{i}\mathcal{D}(  \mathbf{x}_n|\mathbf{w})\right)$. The classification margin is defined as $y_n\mathcal{D}(  \mathbf{x}_n,\mathbf{w})$, and it is large and positive when the label $y_n$ agrees with the prediction. Traditional learning machines such as the max-margin methods learn the optimal parameter setting $\mathbf{w},\mathbf{\gamma}$ by the empirical loss and the regularization penalty as shown below:
\begin{equation}\label{original theory}
\begin{aligned}
&\min_{(\mathbf{w},\mathbf{\gamma}_n)} R(\mathbf{w}) + \sum_{n} L(\mathbf{\gamma}_n)\\
&\quad \text{s.t.} \quad y_n\mathcal{D}(  \mathbf{x}_n\mid\mathbf{w}) - \mathbf{\gamma}_n
 \geq \mathbf{0}, \quad \forall n 
 \end{aligned}
\end{equation}

Where $L()$ is the loss function which is a non-increasing and convex function of the margin, and $R(\mathbf{w})$ is the regularization penalty. However, MED considers a more general problem of finding a distribution $p(\mathbf{w, \gamma})$ over $\mathbf{w}$ and classification margin parameters $\gamma$. This could be done by minimizing its relative entropy with respect to some prior target distribution $p_{0}(\mathbf{w, \gamma})$ under certain margin constraints. Specifically, suppose that a prior distribution, denoted as $p_0(\mathbf{w,\gamma})$, is available, then MED learns a distribution $p(\mathbf{w,\gamma})$ by solving a regularized risk minimization problem. When the prior distribution is not a uniform distribution, this can be generalized as minimizing the relative entropy (or Kullback-Leibler divergence) and the regularization penalty as follows (penalizing larger distances from priors):
\begin{equation}  \label{eq:MED}
 \begin{aligned}
\min _{p\left( \mathbf{w,\gamma}\right) }KL\big( p\left( \mathbf{w,\gamma}\right) ||p_{0}\left(  \mathbf{w,\gamma} \right) \big) + CR\big(p(\mathbf{w,\gamma})\big).
 \end{aligned}
\end{equation}

Here, $C$ is a constant and $R(p\big(\mathbf{w,\gamma})\big) = \sum_n h\big(y_n E_{p(\mathbf{w,\gamma})}[\mathcal{D}\left(   \mathbf{x}_{n}|\mathbf{w} \right)-\mathbf{\gamma}_n]\big)$ is the hinge-loss that captures the large-margin principle underlying the MED prediction rule:
\begin{equation}\label{predrule}
\begin{aligned}
\hat{y} = sign\big(E_{p(\mathbf{w,\gamma})}[\mathcal{D}\left(   \mathbf{x}_{n}|\mathbf{w})-\mathbf{\gamma}_n] \right)\big).
\end{aligned}
\end{equation}

And the KL divergence is defined as follows:
\begin{equation} \label{KL divergence}
\begin{aligned}
KL\big( p\left(  \mathbf{w} ,\mathbf{\gamma} \right) ||p_{0}\left(  \mathbf{w} ,\mathbf{\gamma} \right) \big)=\int p\left(  \mathbf{w} ,\mathbf{\gamma}\right) \log \dfrac {p\left(  \mathbf{w} ,\mathbf{\gamma}\right) }{p_{0} \left(  \mathbf{w, \gamma} \right) }.
\end{aligned}
\end{equation}

Here in \eqref{eq:MED}, the classification margin quantities are included; $\mathbf{\gamma}_n$ as slack variables in the optimization, which represents the minimum margin that $y_n\mathcal{D}(  \mathbf{x}_n|\mathbf{w})$ must satisfy. MED considers an expectation form of the traditional approaches and casts Eq. \eqref{original theory} as an integration. The classification constraints will also be applied in an expected form. As a result, MED no longer finds a fixed set of the parameters, but a distribution over them, and it uses a convex combination of discriminant functions rather than one single discriminant function to make model averaging for decisions. In particular, MED formulation finds distributions that are as close as possible with the prior distribution over all parameters regarding KL-divergence subject to various moment constraints. This analogy extends to cases where the distributions are also over unlabeled samples, missing values, or other probabilistic entities that are introduced when designing the discriminant function. Correspondingly, MED is an effective approach to learn a discriminative classifier as well as consider uncertainties over model parameters, which combines generative and discriminative learning \cite{sun2018multi,zhu2018semi}. This generalization facilitates a number of extensions of the basic approach, including uncertainty quantification described in this paper. The present work contributes by introducing a novel generalization of CHI formulation by integrating the MED to perform the task of uncertainty quantification. 
\section{The proposed work: the UQ-CHI model} \label{proposed work}
The overall goal of UQ-CHI is to learn a distribution
$p(\mathbf{w})$ over the parameters of CHI model $\mathbf{w}$. An additional goal is that this could be done even if only partial labels are given, and the labels might also be with uncertainty. Therefore, the first step in constructing the UQ-CHI is to create the constraint structure. To design the UQ-CHI, we incorporate some features from the original formulation of the CHI via Eq. \eqref{eq:original} as follows: First, we utilize the label information by defining the discriminant function $\mathcal{D}\left(  \mathbf{x}_{n, Tn}|\mathbf{w} \right)=\mathbf{w}^{T}  \mathbf{x}_{n,Tn}$ which corresponds to \eqref{eq:original_b}. We, then incorporate the distinct feature of the CHI formulation, the monotonicity regularization function $\mathcal{M}\left(\mathbf{z}_{n, t}|\mathbf{w} \right)=\mathbf{w}^{T}\mathbf{z}_{n,t}$ that corresponds to Eq. \eqref{eq:original_c}. Note that, here, we will not incorporate the additional terms in Eq. \eqref{eq:original_d} and Eq. \eqref{eq:original_e} as they demand full knowledge of labels of the samples. In addition, we don't include the sparsity regularization term \eqref{eq:original_f}, since our focus is to learn $p(\mathbf{w})$ rather than the parameter vector $\mathbf{w}$. Also, our model can induce sparsity, e.g., if we impose a Laplace prior distribution for the parameters as to what is done in Bayesian Lasso model \cite{park2008bayesian}.

 In the following subsections, we will introduce how we design the prior distributions, the constraints, and how to derive computational algorithms and closed-form solutions for training and prediction. 
\subsection{Design of constraints and prior distributions}\label{constraints}
As aforementioned, there are two types of constraints that we can extract from the CHI formulation into the development of UQ-CHI. One corresponds to the discriminant function $\mathcal{D}\left(  \mathbf{x}_{n, Tn}|\mathbf{w} \right)=\mathbf{w}^{T}  \mathbf{x}_{n,Tn}$ used in CHI, to generate prediction on samples, while the other one corresponds to the monotonicity regularization function $\mathcal{M}\left(\mathbf{z}_{n, t}|\mathbf{w} \right)=\mathbf{w}^{T}\mathbf{z}_{n,t}$. Based on the CHI formulation, it is supposed that the model should lead to $y_n\mathcal{D}\left(  \mathbf{x}_{n, Tn}|\mathbf{w} \right) = 1$ and $\mathcal{M}\left(\mathbf{z}_{n, t}|\mathbf{w} \right)\geq0$. As this perfect model may not exist, a set of margin variables $\mathbf{\gamma}=[\mathbf{\gamma}_1,\ldots,\mathbf{\gamma}_n]$ are introduced. We consider an expectation form of the previous approach and cast Eq. \eqref{eq:original} as an integration. Hence, the classification constraints are applied in an expected sense. This will lead to the following formulation for the constraints:
\begin{subequations} \label{constraintseq}
\begin{align}
&\int p\left( \mathbf{w},\mathbf{\gamma}\right)[p(y_n) \mathcal{D}(  \mathbf{x}_{n,Tn}\mid \mathbf{w}) - \mathbf{\gamma}_{n}]d\mathbf{w}d\mathbf{\gamma} 
+ \label{constraintseq_a} \\
&\int p\left(\mathbf{w,\gamma}\right) \left[\mathcal{M}\left( \mathbf{z}_{n, t}|\mathbf{w} \right) -\mathbf{\gamma}_{n}\right] d\mathbf{w} d\mathbf{\gamma} \geq 0.\label{constraintseq_b}
\end{align}
\end{subequations}

Here, the term \eqref{constraintseq_a} is the discriminant function and the term \eqref{constraintseq_b} is the monotonicity regularization function. And, $p(y_n)$ is the distribution of $y_n$, and $p(\mathbf{w,\gamma})$ is the distribution of $\mathbf{w,\gamma}$. With the prior distribution, we can derive the prediction rule: $\hat{y} = sign(E_{p(\mathbf{w})}[\mathcal{D}\left(  \mathbf{x}_{n}|\mathbf{w}] \right))$. 

Now we move on to the design of the prior distribution $p_0(\mathbf{w,\gamma}, y)$. It is natural to decompose the joint prior distribution as a product of three distributions:
\begin{equation}  \label{eq:PriorMEDCHI}
 \begin{aligned}
p_0(\mathbf{w,\gamma}, y) = p_0(\mathbf{w})\prod ^{N}_{1}p_{0}\left( \mathbf{\gamma}_n\right)\prod ^{N}_{1}p_{0}\left( y_n\right).
 \end{aligned}
\end{equation}

In what follows we discuss each of the three prior distributions. Specifically, it is reasonable to assume that a level of uncertainty can be designed to each example in defining $p_0(y_n)$. A simple solution is to set $p_{0}\left( y_n\right) =1$ whenever $y_n$ is observed and $p_{0}\left( y_n\right) =0.5$ otherwise. To define $p_{0}\left(\mathbf{w}\right)$, we choose $p_{0}\left(\mathbf{w}\right)$ to be a Gaussian distribution with mean vector as $\mathbf{0}$ and covariance matrix as an identity matrix $\mathbf{I}$. To define the prior over the margin variables, we assume that it could be factorized $p\left(\mathbf{\gamma}\right) =\prod _{n}p_{0}\left(\mathbf{\gamma}_n\right)$. Further, following the idea proposed in \cite{jaakkola2000maximum}, we can set $p_{0}\left( \mathbf{\gamma}_n\right)= ce^{-c\left( 1-\mathbf{\gamma}_{n}\right) }$ and $\mathbf{\gamma}_{n}\leq 1$. Here, $1-\frac{1}{c}$ is actually the mean of the prior distribution of $\mathbf{\gamma}_n$, so the idea of this distribution is to incur a penalty only for margins smaller than $1-\frac{1}{c}$, while for margins larger than this quantity are not penalized. More details about the design of prior distributions will be given in Section \ref{tract}.
\subsection{The computational algorithm for UQ-CHI}\label{UQ-CHI solution}
The full formulation of the proposed UQ-CHI model is shown below:
\begin{subequations}\label{full UQ-CHI}
\begin{align}
&\min _{p\left( \mathbf{w,\gamma}\right) }KL\big( p\left( \mathbf{w,\gamma}\right) ||p_{0}\left(  \mathbf{w,\gamma} \right) \big)\label{full UQ-CHI_a}\\
 &\quad \textrm{s.t.}  \int p\left( \mathbf{w},\mathbf{\gamma}\right)[p(y_n) \mathcal{D}(  \mathbf{x}_{n,Tn}\mid \mathbf{w}) - \mathbf{\gamma}_{n}]d\mathbf{w}d\mathbf{\gamma} + \label{full UQ-CHI_b} \\
&\qquad \int p\left(\mathbf{w,\gamma}\right) \left[\mathcal{M}\left( \mathbf{z}_{n, t}|\mathbf{w} \right) -\mathbf{\gamma}_{n}\right] d\mathbf{w} d\mathbf{\gamma} \geq 0.\label{full UQ-CHI_c}
\end{align}
\end{subequations}

Essentially, solving optimization formulation Eq. \eqref{full UQ-CHI} is to find a solution by calculating the relative entropy projection from the overall prior distribution $p_0(\mathbf{w, \gamma}, y)$ to the admissible set of distributions $p$ that are consistent with the constraints. In what follows, we develop the computational algorithm to solve this formulation Eq. \eqref{full UQ-CHI} and further derive the method for the prediction on samples. 
\subsubsection{Step 1: Training the model}
In the training step, we consider a joint distribution of $\mathbf{w}$, and the margin vector of $\mathbf{\gamma}=[\mathbf{\gamma}_1,\ldots,\mathbf{\gamma}_n]$ while fixing $p(y_n)$.  In this step, we first explain the solution to the MED optimization problem subject to the terms in \eqref{eq:MED}. 
\begin{lem} \label{lemma_1}
Let the loss function be a non-increasing and convex function of the margin, and let the Lagrangian of the optimization problem defined as $\mathcal{L}$ and  $\mathbf{\lambda}=\left[ \mathbf{\lambda} _{1},\ldots ,\mathbf{\lambda} _{n}\right]$ be a set of non-negative Lagrange multipliers. Given the prior distribution $p_0(\mathbf{w})$ and the model distribution $p(\mathbf{w})$, and the discriminant function $\mathcal{D}\left(   \mathbf{x}_{n}|\mathbf{w} \right)$ in order to minimize the relative entropy in terms of the KL-divergence ($KL(p(\mathbf{w}||p_0(\mathbf{w})$) subjected to set of defined constraints, the MED optimization problem \eqref{eq:MED} can be written as: 
\begin{equation}\label{regularZ}
\begin{aligned}
 & \max _{\lambda }J(\mathbf{\lambda}) = -logZ(\mathbf{\lambda})\\
 &\quad \textrm{s.t.} \quad \mathbf{\lambda}_i \geq 0 \quad \text{for} \quad i=1,\ldots,N
\end{aligned}
\end{equation}
Here, $Z(\mathbf{\lambda})$ is the normalization constant defined as:
\begin{equation}\label{lemma_1_Z}
\begin{aligned}
&Z(\mathbf{\lambda})=\int p_0(\mathbf{w}) \exp \Bigg(\sum_n \mathbf{\lambda}_ny_n \mathcal{D}\left(  \mathbf{x}_{n}|\mathbf{w} \right)\Bigg)d\mathbf{w},
\end{aligned}
\end{equation}
\end{lem} 

The proof of Lemma \ref{lemma_1} can be found in \ref{Appendix_A}. Now, the model training problem is revealed to be another optimization problem, that is learning optimal $\mathbf{\lambda}^*$ by solving the dual objective function $J$ under positivity constraint. Based on the results from Lemma \ref{lemma_1}, after adding dual variables for the constraint in Eq. \eqref{full UQ-CHI}, the Lagrangian of the optimization problem can be written as:
\begin{equation}  \label{eq:MMED-CHIlagrang}
 \begin{aligned}
&\mathcal{L}  = \int p\left(\mathbf{w,\gamma}\right) \log \dfrac {p\left( \mathbf{w,\gamma}\right) }{p_{0}\left( \mathbf{w,\gamma}\right) }d\mathbf{w} d\mathbf{\gamma}-\\
 &\qquad \Bigg(\sum _{n\in \left\{ 1,\ldots,N\right\} }\int p\left( \mathbf{w},\mathbf{\gamma}\right)\mathbf{\lambda}_n[p(y_n) \mathcal{D}(  \mathbf{x}_{n,Tn}\mid \mathbf{w}) - \mathbf{\gamma}_{n}]d\mathbf{w}d\mathbf{\gamma} + \\
&\quad \sum_{\substack{n\in \{1,\ldots,N\}\\ t\in \left\{ 1,\ldots ,T_{n-1}\right\}} }\int p\left(\mathbf{w,\gamma}\right)\mathbf{\lambda}_n \left[\mathcal{M}\left( \mathbf{z}_{n, t}|\mathbf{w} \right) -\mathbf{\gamma}_{n}\right] d\mathbf{w} d\mathbf{\gamma}\Bigg).
 \end{aligned}
\end{equation}
In order to find a solution, we require:
\begin{equation}\label{MMEDCHIDerivative}
\begin{aligned}
&\dfrac {\partial \mathcal{L}}{\partial p\left(\mathbf{w,\gamma} \right) }  = \log \dfrac {p\left(\mathbf{w,\gamma}\right) }{p_{0}\left(\mathbf{w,\gamma}\right)}+1 -\\
&\qquad \qquad \quad  \Bigg ( \sum _{n\in \left\{ 1,\ldots,N\right\} }\mathbf{\lambda}_n[p(y_n) \mathcal{D}(  \mathbf{x}_{n,Tn}\mid \mathbf{w}) - \mathbf{\gamma}_{n}]  + \\
 &\qquad \quad \quad \sum_{\substack{n\in \{1,\ldots,N\}\\ t\in \left\{ 1,\ldots ,T_{n-1}\right\}} }\mathbf{\lambda}_n \left[\mathcal{M}\left( \mathbf{z}_{n, t}|\mathbf{w} \right) -\mathbf{\gamma}_{n}\right]\Bigg)\\
 &\qquad \qquad = \mathbf{0},
\end{aligned}
\end{equation}
Which results in the following theorem. 
\begin{theorem}\label{theory_1}
The solution to the UQ-CHI problem has the following general form:
\begin{equation}  \label{eq:MMEDCHIsol}
 \begin{aligned}
&p\left(\mathbf{w,\gamma}\right)^* =  \frac{1}{Z(\mathbf{\lambda})}p_{0}\left( \mathbf{w,\gamma}\right)\\
&\qquad \quad \quad \exp \Bigg (\sum _{n\in \left\{ 1,\ldots ,N\right\} }\mathbf{\lambda}_n[p(y_n) \mathcal{D}(  \mathbf{x}_{n,Tn}\mid \mathbf{w}) - \mathbf{\gamma}_{n}]+ \\ 
&\qquad \quad\sum_{\substack{n\in \{1,\ldots,N\}\\ t\in \left\{ 1,\ldots ,T_{n-1}\right\}} } \mathbf{\lambda}_n\left[\mathcal{M}\left( \mathbf{z}_{n, t}|\mathbf{w} \right) -\mathbf{\gamma}_{n}\right] \Bigg) .
 \end{aligned}
\end{equation}
\end{theorem}
Thus, finding the solution to \eqref{full UQ-CHI} depends on being able to evaluate the normalization constant $Z(\mathbf{\lambda})$.
\begin{lem} \label{lemma_2}
Let $Z(\mathbf{\lambda})$ be the normalization constant defined in Eq. \eqref{lemma_1_Z}. Based on the finding in \eqref{eq:MMEDCHIsol}, $Z(\mathbf{\lambda})$ can be reformulated as follows:
\begin{subequations}\label{finalZZ}
\begin{align}
&Z(\mathbf{\lambda})= Z_\mathbf{w}(\mathbf{\lambda}) + Z_\mathbf{\gamma}(\mathbf{\mathbf{\lambda}})\\
&\qquad =\exp \Bigg(\frac{1}{2}\bigg(\sum_{\substack{n\in \{1,\ldots,N\}} }\mathbf{\lambda}_{n} p_n(y_n)  \mathbf{x}_{n,Tn} \sum_{\substack{n\in \{1,\ldots,N\}\\ t\in \left\{ 1,\ldots ,T_{n-1}\right\}} } \mathbf{\lambda}_n \mathbf{z}_{n,t}\bigg )^T\label{finalZZ_a}\\
&\qquad \quad \bigg(\sum_{\substack{n\in \{1,\ldots,N\}} }\mathbf{\lambda}_{n} p_n(y_n)  \mathbf{x}_{n,Tn} \sum_{\substack{n\in \{1,\ldots,N\}\\ t\in \left\{ 1,\ldots ,T_{n-1}\right\}} } \mathbf{\lambda}_n \mathbf{z}_{n,t}\bigg)\Bigg)+ \label{finalZZ_b}\\
&\quad \quad\prod_{n\in \{1,\ldots,N\} }\frac{1}{1-\mathbf{\lambda}_n/c}\exp(-\mathbf{\lambda}_n)\label{finalZZ_c}.
 \end{align}
\end{subequations}
Where, $Z_\mathbf{w}(\mathbf{\lambda})$ is defined in \eqref{finalZZ_a} and \eqref{finalZZ_b} $Z_\mathbf{\gamma}(\mathbf{\lambda})$ is defined in \eqref{finalZZ_c}.
\end{lem}

The proof of Lemma \ref{lemma_2} can be found in the \ref{Appendix_B}. Given the reformulated normalization constant $Z(\mathbf{\lambda})$ in \eqref{finalZZ}, the maximum of the jointly concave function objective function $J(\mathbf{\lambda})$ showing in Eq. \eqref{regularZ} can be found through a constrained non-linear optimization. As a result, by substituting Eq. \eqref{finalZZ} in Eq. \eqref{regularZ} we get:
\begin{equation}\label{seven}
\begin{split} 
&J(\mathbf{\lambda})=\sum _{n\in \{1,\ldots,N\}}\bigg( \mathbf{\mathbf{\lambda}}_{n}+log (1-\mathbf{\lambda}_{n}/c)\bigg)-\\
&\qquad \qquad\frac{1}{2}\bigg(\sum_{\substack{n\in \{1,\ldots,N\}} }\mathbf{\lambda}_{n} p_n(y_n)  \mathbf{x}_{n,Tn} \sum_{\substack{n\in \{1,\ldots,N\}\\ t\in \left\{ 1,\ldots ,T_{n-1}\right\}} } \mathbf{\lambda}_n \mathbf{z}_{n,t}  \bigg)^T \\
 & \qquad\qquad \bigg(\sum_{\substack{n\in \{1,\ldots,N\}} }\mathbf{\lambda}_{n} p_n(y_n)  \mathbf{x}_{n,Tn} \sum_{\substack{n\in \{1,\ldots,N\}\\ t\in \left\{ 1,\ldots ,T_{n-1}\right\}} } \mathbf{\lambda}_n \mathbf{z}_{n,t}\bigg).
\end{split}
\end{equation}

Here, $\lambda \geq \mathbf{0}$. Thus, we have the following dual optimization problem:
\begin{equation}\label{MMEDCHIdual}
\begin{aligned}
&\max _{\mathbf{\lambda}}\sum _{n\in \{1,\ldots,N\}}\bigg( \mathbf{\lambda}_{n}+log (1-\mathbf{\lambda}_{n}/c)\bigg) -\\
&\qquad \quad \frac{1}{2}\bigg(\sum_{\substack{n\in \{1,\ldots,N\}} }\mathbf{\lambda}_{n} p_n(y_n)  \mathbf{x}_{n,Tn} \sum_{\substack{n\in \{1,\ldots,N\}\\ t\in \left\{ 1,\ldots ,T_{n-1}\right\}} } \mathbf{\lambda}_n \mathbf{z}_{n,t}  \bigg)^T \\
&\qquad \quad \bigg(\sum_{\substack{n\in \{1,\ldots,N\}} }\mathbf{\lambda}_{n} p_n(y_n)  \mathbf{x}_{n,Tn} \sum_{\substack{n\in \{1,\ldots,N\}\\ t\in \left\{ 1,\ldots ,T_{n-1}\right\}} } \mathbf{\lambda}_n \mathbf{z}_{n,t}\bigg) \\
&\quad \textrm{s.t.} \quad  \mathbf{\lambda} \geq \mathbf{0}
\end{aligned}
\end{equation}

The Lagrange multiplier $\mathbf{\lambda}$, is recovered by solving the convex optimization problem Eq. \eqref{MMEDCHIdual}. Note that since the prior factorizes across $\mathbf{w,\gamma}$, UQ-CHI solution also factorized as well, i.e., $p(\mathbf{w,\gamma})=p(\mathbf{w})p(\mathbf{\gamma})$. 
\begin{cor}\label{cor_1}
From results in Theorem \ref{theory_1} the marginal distribution $p(\mathbf{w})$ can be found as follows:
\begin{equation}\label{distributionestimate}
\begin{aligned}
& p\left(\mathbf{w}\right)=\frac{1}{Z_\mathbf{w}(\mathbf{\lambda})}
p_{0}\left( \mathbf{w}\right) \exp \bigg(\sum_{\substack{n\in \{1,\ldots,N\}}}\mathbf{\lambda}_n p_n(y_n)\mathbf{w}^T  \mathbf{x}_{n,Tn}+ \\
&\qquad \sum _{\substack{n\in \{1,\ldots,N\}\\ t\in \left\{ 1,\ldots ,T_{n-1}\right\}}}\mathbf{\lambda}_{n} \mathbf{w}^T\mathbf{z}_{n, t}\bigg). 
\end{aligned}
\end{equation}
\end{cor}

Where, $Z_{\mathbf{w}}(\mathbf{\lambda})$  can be obtained from Eq. \eqref{finalZZ_a} and \eqref{finalZZ_b}.
\subsubsection{Step 2: Prediction}\label{prediction}
After obtaining the marginal distribution $p(\mathbf{w})$ in \eqref{distributionestimate}, the following lemma is used to predict the label of a new example $  \mathbf{x}_{new}$. Referring to the solution of the UQ-CHI problem in \eqref{eq:MMEDCHIsol}, we can easily modify the regularization approach for predicting a new label from a new input sample $  \mathbf{x}_{new}$ that is shown by $\hat{y}=\textit{sign } \mathcal{D}\left(   \mathbf{x}| \mathbf{\hat w} \right)$. In what follows, we generate the predictive label for the upcoming new labels.
\begin{lem}\label{lemma_3}
Given the marginal distribution $p\left(\mathbf{w}\right)$ in \eqref{distributionestimate}  and the convex combination of discriminant functions $\int p(\mathbf{w})\mathcal{D} \left(  \mathbf{x}|\mathbf{w}\right)d\mathbf{w}$, and let $\mathbf{\lambda}^*$ be the optimal Lagrangian multiplier obtained from the optimization problem \eqref{MMEDCHIdual}, and given $Z_\mathbf{w}(\mathbf{\lambda})$ obtained from \eqref{finalZZ_c}, then the predictive label for the new ($  \mathbf{x}_{new}$) can be generated as:
\begin{equation}\label{predict}
\begin{aligned}
&\hat {y}=sign \Bigg(\bigg(\sum_{\substack{n\in \{1,\ldots,N\}}}\mathbf{\lambda}_n p_n(y_n)  \mathbf{x}_{n,Tn} + \sum _{\substack{n\in \{1,\ldots,N\}\\ t\in \left\{ 1,\ldots ,T_{n-1}\right\}}}\mathbf{\lambda}_{n} \mathbf{z}_{n, t}\bigg)^T  \mathbf{x}_{new}\Bigg). 
\end{aligned}
\end{equation}
\end{lem}
The proof of Lemma \ref{lemma_3} is shown in \ref{Appendix_C}. 

\subsubsection{Summary of the algorithms}\label{summary}
A full description of the training and prediction of UQ-CHI model is given in Algorithm \ref{alg:alg1}. 

\begin{algorithm}[!ht]                    
\caption{The UQ-CHI algorithm~}           
\label{alg:alg1}                          
\begin{algorithmic}[1]                    
\Require   $  \mathbf{x}_{n, t} \in \mathbb{R} ^{d}$, $p_{0}\left( y_n\right)$, $p_{0}\left( \mathbf{w}\right)$, and $p_{0}\left( \mathbf{\gamma}\right)$
 \Ensure Generate predictive labels for the upcoming new labels
\While {not converge}
\State \textbf{Start iterations } t:= 1,2,\ldots\textbf{do}
\State   \textbf{Step 1 - Training model: find $\mathbf{\lambda}^*$ and $p(\mathbf{w})$}
\State   \textbf{for} $n=1,2,\ldots ,N$,
\State    $\max _{\mathbf{\lambda}}\sum _{n\in \{1,\ldots,N\}}\bigg( \mathbf{\lambda}_{n}+log (1-\mathbf{\lambda}_{n}/c)\bigg) -$\\
$\qquad \quad \frac{1}{2}\bigg(\sum_{\substack{n\in \{1,\ldots,N\}} }\mathbf{\lambda}_{n} p_n(y_n)  \mathbf{x}_{n,Tn} \sum_{\substack{n\in \{1,\ldots,N\}\\ t\in \left\{ 1,\ldots ,T_{n-1}\right\}} } \mathbf{\lambda}_n \mathbf{z}_{n,t}  \bigg)^T$ \\
$\qquad \quad \bigg(\sum_{\substack{n\in \{1,\ldots,N\}} }\mathbf{\lambda}_{n} p_n(y_n)  \mathbf{x}_{n,Tn} \sum_{\substack{n\in \{1,\ldots,N\}\\ t\in \left\{ 1,\ldots ,T_{n-1}\right\}} } \mathbf{\lambda}_n \mathbf{z}_{n,t}\bigg)$ \\
$\quad \textrm{s.t.} \quad  \mathbf{\lambda} \geq \mathbf{0}$
\State $ p\left(\mathbf{w}\right)=\frac{1}{Z_\mathbf{w}(\mathbf{\lambda})}
p_{0}\left( \mathbf{w}\right) \exp \bigg(\sum_{\substack{n\in \{1,\ldots,N\}}}\mathbf{\lambda}_n p_n(y_n)\mathbf{w}^T  \mathbf{x}_{n,Tn}+ $\\
$\qquad \sum _{\substack{n\in \{1,\ldots,N\}\\ t\in \left\{ 1,\ldots ,T_{n-1}\right\}}}\mathbf{\lambda}_{n} \mathbf{w}^T\mathbf{z}_{n, t}\bigg) $
\State   \textbf{Step 2 - Prediction: predict the
label of a new example ($  \mathbf{x}_{new}$)}
\State ${\displaystyle\hat {y}=sign \Bigg(\bigg(\sum_{\substack{n\in \{1,\ldots,N\}}}\mathbf{\lambda}_n p_n(y_n)  \mathbf{x}_{n,Tn} + \sum _{\substack{n\in \{1,\ldots,N\}\\ t\in \left\{ 1,\ldots ,T_{n-1}\right\}}}\mathbf{\lambda}_{n} \mathbf{z}_{n, t}\bigg)^T   \mathbf{x}_{new}\Bigg)}$
\State   \textbf{ end for}
\EndWhile
\end{algorithmic}
\end{algorithm}
\subsection{UQ-CHI with rejection option}\label{rejction option}
Typically the performance of a prediction model is evaluated based on its accuracy, on a scheme of classifying all samples, regardless of the degree of confidence associated with the classification of the samples. However, accuracy is not the only measurement that can be used to judge the model’s performance.  In many healthcare application, it is safer to make predictions when the confidence assigned to the classification is relatively high, rather than classify all samples even if confidence is low. In this case, a sample can be rejected if it doesn't fit into any of the classes. In pattern recognition, this problem is typically solved by estimating the class conditional probabilities and rejecting the samples that have the lowest class posterior probabilities, that are the most unreliable samples. As UQ-CHI enables uncertainty quantification, here, we create a rejection option in prediction to show the utility of uncertainty quantification in practice. The basic idea of rejection option is that the prediction model rejects to generate a prediction if the uncertainty is higher than a given threshold.  In other words, a sample that is most likely to be misclassified is rejected as described  below:
\begin{equation}
\begin{aligned}
\max  p(\mathbf{w}|x_i) < T \quad i = 1,\dots,N
\end{aligned}
\end{equation}

Here, T is the rejection rate. The samples $x_i$ are rejected for which the maximum posterior probability $p(\mathbf{w}|x_i)$ is below a threshold. And a sample is accepted when:
\begin{equation}
\begin{aligned}
\max  p(\mathbf{w}|x_i) \geq T \quad i = 1,\dots,N
\end{aligned}
\end{equation}

Thus, we define a classification with rejection as $ \hat {y}^{rejection}$, where, if a sample is rejected $\hat {y}^{rejection}_i = 0$, denotes rejection, else,  $\hat {y}^{rejection}_i = \hat {y}_i$,  where, $\hat {y}_i$ corresponds to the classification of the $i$th sample defined in Eq.  \eqref{predict}.
\begin{table*}[t]
\centering
\resizebox{0.60\textwidth}{!}{%
\begin{tabular}{|l|l|l|l|l|l|}
\hline
\multicolumn{1}{|c|}{Algorithm name} & \multicolumn{4}{c|}{UQ-CHI} & \multicolumn{1}{c|}{\multirow{3}{*}{CHI}} \\ \cline{1-5}
\multicolumn{1}{|c|}{\multirow{2}{*}{Label ratio}} & \multicolumn{1}{c|}{\multirow{2}{*}{Training ratio}} & \multicolumn{3}{c|}{Rejection rate} & \multicolumn{1}{c|}{} \\ \cline{3-5}
\multicolumn{1}{|c|}{} & \multicolumn{1}{c|}{} & Low = 20 & Medium = 40 & High = 60 & \multicolumn{1}{c|}{} \\ \hline
\multirow{3}{*}{Low = 10} & 30 & 0.69 & 0.74 & 0.81 & 0.61 \\ \cline{2-6} 
 & 50 & 0.73 & 0.76 & 0.83 & 0.62 \\ \cline{2-6} 
 & 70 & 0.75 & 0.77 & 0.85 & 0.65 \\ \hline
\multirow{3}{*}{Medium = 20} & 30 & 0.66 & 0.72 & 0.73 & 0.55 \\ \cline{2-6} 
 & 50 & 0.69 & 0.73 & 0.74 & 0.60 \\ \cline{2-6} 
 & 70 & 0.71 & 0.75 & 0.78 & 0.64 \\ \hline
\multirow{3}{*}{High = 50} & 30 & 0.64 & 0.69 & 0.72 & 0.53 \\ \cline{2-6} 
 & 50 & 0.67 & 0.71 & 0.73 & 0.56 \\ \cline{2-6} 
 & 70 & 0.70 & 0.73 & 0.75 & 0.60 \\ \hline
\end{tabular}}
\caption{Corresponding testing accuracies for different rejection options for the simulated dataset}
\label{Table_3}
\end{table*}
\subsection{Tractability of UQ-CHI related to design of prior distribution}\label{tract}
Recall that by applying the MED to our optimization problem we no longer learn the model parameter, and instead, we specify the probability distributions. These distributions give rise to penalty functions for the model and the margins via KL-divergence. In detail, the model distribution will give rise to a divergence term $KL(p(\mathbf{w})||p_0(\mathbf{w}))$, and the margin distribution will give rise to the divergence term $KL(p(\mathbf{\gamma})||p_0(\mathbf{\gamma}))$ which corresponds to the regularization penalty and the loss function respectively. The trade-off between classification loss and regularization now are on a common probabilistic scale, since both terms are based on probability distributions and KL-divergence. Hence, there is a  relationship between defining a prior distribution over margins and parameters and defining the objective function and the penalty term in the original function. Recall that, $\mathbf{\gamma}_n$ are the classification margins as slack variables in the optimization which represent the minimum margin that $y_n\mathcal{D}(X_n;w)$ must satisfy. Hence, the choice of the margin distribution corresponds to the use of the slack variables in the formulation of the UQ-CHI. For example, in our case we set $p_{0}\left( \mathbf{\gamma}_n\right)= ce^{-c\left( 1-\mathbf{\gamma}_{n}\right) }$ and $\mathbf{\gamma}_{n}\leq 1$. If we mathematically expand the normalization function in \eqref{lemma_1_Z}, we get the two terms $Z_{\mathbf{w}}(\mathbf{\lambda})$ and $Z_{\mathbf{\gamma}}(\mathbf{\lambda})$ as shown in \eqref{finalZZ}, and given the choice of margin priors in Section \ref{constraints} we get:
\begin{equation}\label{penalty}
\begin{aligned}
log Z_{\mathbf{\gamma}_n}(\mathbf{\lambda}_n) =log \int ^{1}_{\mathbf{\gamma}_n = -\infty} c\exp\bigg({-c\big( 1-\mathbf{\gamma}_{n}\big) }\bigg)\exp\bigg(-\mathbf{\lambda}_n\mathbf{\gamma}_n\bigg)d\mathbf{\gamma}_n.
\end{aligned}
\end{equation}

From \eqref{penalty} we can see that a penalty occurs when the margins are smaller than $E[\mathbf{\gamma}_n] = 1 - \frac{1}{c}$, and any margins larger than this would not be penalized. The margin distribution becomes peaked when $\mathbb{\gamma}_n = 1$ that is when $c\rightarrow \infty$, and this is equivalent to having fixed margins. If the margin values are held fixed the discriminant function might not be able to separate the training examples with such pre-specified margin values. Because of non-separable datasets this will generate an empty convex hull for the solution space. Thus, we need to revisit the setting of the margin values, and the loss function upon them. The parameter $c$ will play an almost identical role as the regularization parameter which upper bounds the Lagrange multipliers. Note, if the objective function $J()$ grow without a bound, it may generate a search space for parameters that are no longer a convex hull. This compromises the uniqueness and solvability of the problem. Therefore, the selection of a prior forms a concave function $J()$ for a unique optimum in the Lagrange multiplier space. 
\section{Numerical studies} \label{numerical studies}
In this section, we design our simulation studies to evaluate the efficacy of UQ-CHI in terms of prediction and uncertainty quantification, in comparison with the CHI model under a variety of practical scenarios. 
\subsection{Simulated dataset} \label{simulated dataset}
We simulate data following the procedure described as follows. The synthetic dataset is generated with two classes with partial labels. We conduct several experiments with the simulated data to investigate the performance of our method across different settings.  Without loss of generality, we assume that there are two groups, normal vs. diseased with a proportion of $60\%$ of class normal and $20\%$ of complete labels. For all the experiments, we set the number of features $d= 90$, For each class, we simulate $50$ subjects, where we assumed that $x^{k}_{n,t}\sim
N\big( u,\sigma ^{2}_{k}\big)$ for $k\in \left\{ 1,\ldots ,d\right\}$.
\subsection{Incomplete labels and length of longitudinal data}
UQ-CHI can handle partial labels well, i.e., by assigning a prior distribution of the labels and obtaining posterior distributions after model training, in our experiment, we consider a low, medium and high level of label availability, i.e., $10\%$, $20\%$ and $50\%$ of unlabeled examples. Also, we evaluate our methodology's robustness in the presence of down-sampling of the training data, i.e., only using a percentage of the data (for example, ranging from $30\%$, $50\%$ and  $70\%$), to train both UQ-CHI and CHI models. A model that can predict well with less longitudinal data holds great value in clinical applications. 
\subsection{Uncertainty quantification with rejection option}
As mentioned in \ref{rejction option}, UQ-CHI has a unique capacity of rejection option. The algorithm rejects to predict on a sample if it cannot be predicted reliably. The key parameter is the threshold that will be used in the rejection option. In our experiments, we use several levels of the threshold to create a range of rejection options from loose to strict, and further calculate the resulting accuracies on the predictions on the accepted samples. Specifically, we vary the size of the rejection region from $20\%$, $40\%$, to $60\%$. 
\subsection{Parameter tuning and validation }
In our experiments, we randomly split the data into two parts, one for training and one for testing. For the training dataset, we use 10-fold cross-validation to tune the parameters.  The average accuracies from the split of the testing dataset are reported in the result section. In Section \ref{tract} we specify under what condition the computation would remain tractable. It has been pointed out that, based on the choice of the margin distribution described in \ref{tract}, $\mathbf{\gamma}_n$ is bounded by the parameter $c$. Recall that $c$ is a parameter in the prior for the margins. Therefore, the parameter $c$ will play an important role. Hence, we conduct experiments with the parameter $c$ chosen from $\left(1.5,3,5,10,20,100 \right)$ to see the impact of various choices of $c$ on the testing accuracy.
\subsection{Discussion}\label{Discussion}
In the following, we discuss the tractability of the model given the simulated data for various choices of the parameter $c$ in Table \ref{Table_1}. We simulated different selection of the parameter $c$ to check its impact on the testing accuracy. If we observe that increasing this parameter imposes no effect on the performance, we would then ignore the higher values for reasons discussed in Section \ref{tract}. The results show that for a more significant quantity of parameter $c$ the accuracy decreases. As shown in Table \ref{Table_1}, additional potential terms of the parameter $c$ would not carry huge effects as the margin distribution may have become at its peak ($\mathbf{\gamma}_n$) which is equivalent to have fixed margins. Note that to test the impact of the parameter $c$ we simulated the data with a proportion of $60\%$ of class normal and $20\%$ complete labels. Here we can observe that after increasing the values for the parameter $c$ beyond $5$, the performance of the model doesn't change significantly, which indicates that the margin distribution may have become at its peak, and hence it is equal to a fixed value. Higher values of this parameter generate relatively similar performance. Consequently, lower values of $c$ preserve flexibility to estimate a distribution over parameters instead of using fixed margins.

Next, we examine how the incomplete label information would affect the performance of UQ-CHI with regards to the testing accuracy given different sampling ratios in Table \ref{Table_2}. A model which can be trained with less training data is more promising in healthcare applications where the data collection is relatively costlier than other real-world applications. The results in Table \ref{Table_2} show that with even a ratio of $50\%$ of incomplete label information the UQ-CHI can perform with a testing accuracy of $74\%$. This confirms that the model is capable of performing well in the face of lack of label information. 

Incorporating a rejection option into the model improves the prediction accuracy of classifiers. There is a general relationship between the testing accuracy and rejection rate: the testing accuracy increases monotonically with increasing rejection rate. The testing accuracies for different rejection options are reported in Table \ref{Table_3}. Comparisons of varying rejection rates for the UQ-CHI confirms that for a high rejection rate of $60\%$, the testing accuracy could go up to $81\%$ for a given label ratio of $10\%$, which in comparison with a lower rejection rate, this can be a promising result. In Table \ref{Table_3}, we also compared our methodology with CHI framework. Recall that CHI is not strictly a supervised learning problem. In \cite{huang2017chi},
both simulation studies and real-world applications demonstrated that without label information, CHI method could still be trained and used to predict. However, we show that the UQ-CHI can generate relativity a better performance than CHI by incorporating the rejection option. UQ-CHI can obtain a testing accuracy in a range of $75\%$ to $81\%$ for a given rejection rate of $60\%$ and a labeling ratio of $10\% - 50 \%$. 
\begin{table}[!ht]
\centering
\resizebox{0.23\textwidth}{!}{%
\begin{tabular}{|l|l|}
\hline
Parameter c & Testing accuracy \\ \hline
1.5 & 81.2 \\ \hline
3 & 80.2 \\ \hline
5 & 79.8 \\ \hline
10 & 77.2 \\ \hline
20 & 77.3 \\ \hline
100 & 76.1 \\ \hline
\end{tabular}}
\caption{Model average testing accuracy ($\%$) for simulated dataset}
\label{Table_1}
\end{table}
\begin{table}[t]
\centering
\resizebox{0.45\textwidth}{!}{%
\begin{tabular}{|l|l|l|l|}
\hline
\multicolumn{1}{|c|}{\multirow{2}{*}{Sample ratio}} & \multicolumn{3}{c|}{Label ratio} \\ \cline{2-4} 
\multicolumn{1}{|c|}{} & Low = 10\% & Medium = 20\% & High = 50\% \\ \hline
30 & 0.85 $\pm$ 0.033 & 0.80 $\pm$ 0.032 & 0.74 $\pm$ 0.033 \\ \hline
50 & 0.86 $\pm$ 0.060 & 0.83 $\pm$ 0.053 & 0.76 $\pm$ 0.027 \\ \hline
70 & 0.88 $\pm$ 0.074 & 0.85 $\pm$ 0.041 & 0.78 $\pm$ 0.037 \\ \hline
\end{tabular}}
\caption{The average classification accuracies and standard deviations (\%) for the simulated dataset}
\label{Table_2}
\end{table}
\section{Real-world application on Alzheimer's disease} \label{ADNI}
We further test UQ-CHI on an Alzheimer's disease data which exhibited monotonic disease progression. We use the FDG-PET images of 162 patients (Alzheimer's Disease: 74, Normal aging: 88) downloaded from the ADNI (\url{www.loni.usc.edu/ADNI}). The data is sampled at irregular time points where each patient has at least three time points and at most seven-time points. The data is preprocessed, and the Automated Anatomical Labeling (AAL) is used to segment each image into 116 anatomical volumes of interest (AVOIs). For this study,  90 AVOIs that are in the cerebral cortex are selected (each AVOI becomes a variable here). According to the mechanism of FDG-PET, the measurement data of each region are the local average FDG binding counts, which represents the degree of glucose metabolism.  The glucose metabolism declines as the function of aging, and the progression of many neurodegenerative diseases such as AD further accelerates this declination. Thus, ADNI dataset facilitates a perfect application example to test the proposed method. While the ADNI dataset consists of fully labeled examples, we exploit the dataset settings to create a variety of uncertainties to the label information.

The results for tuning the parameter $c$ for the ADNI dataset is reported in Table \ref{Table_4}. The results show that for a more significant quantity of parameter $c$ the accuracy decreases.   Table \ref{Table_5} shows the performance of the UQ-CHI across different uncertainty levels as well as different sampling ratios. The proposed method shows an excellent capability to quantify the uncertainties for the real-world dataset. As shown in Table \ref{Table_5}, The UQ-CHI is even capable of dealing with a data that has $50\%$ of incomplete labels with an accuracy in the range of $70\%-77\%$ for the ADNI dataset.

On the other hand, we show that by only using a small proportion of the training samples as low as $30\%$ of the data, we still can maintain reasonable performance in a range of $70\%-82\%$, which indicates that UQ-CHI can be trained with less training data. The rejection options against the testing accuracy as well as these values against the training ratios are shown in Tables \ref{Table_6}.   Incorporating a rejection option into the model improves the prediction accuracy of classifiers. Comparisons of different rejection rates for the UQ-CHI confirms that for a high rejection rate of $60\%$, the testing accuracy could go up to $80\%$ or higher, which compared with a lower rejection rate, this can be a promising result. 
\begin{table}
\centering
\resizebox{0.23\textwidth}{!}{%
\begin{tabular}{|l|l|}
\hline
Parameter c & Testing accuracy \\ \hline
1.5 & 78.8 \\ \hline
3 & 77.9 \\ \hline
5 & 77.3 \\ \hline
10 & 75.3 \\ \hline
20 & 72 \\ \hline
100 & 68.9 \\ \hline
\end{tabular}}
\caption{Model average testing accuracy ($\%$) for ADNI dataset}
\label{Table_4}
\end{table}
\begin{table}
\centering
\resizebox{0.45\textwidth}{!}{%
\begin{tabular}{|l|l|l|l|}
\hline
\multicolumn{1}{|c|}{\multirow{2}{*}{Sample ratio}} & \multicolumn{3}{c|}{Label ratio} \\ \cline{2-4} 
\multicolumn{1}{|c|}{} & Low = 10\% & Medium = 20\% & High = 50\% \\ \hline
30 & 0.82 $\pm$ 0.022 & 0.79 $\pm$ 0.052 & 0.70 $\pm$ 0.032 \\ \hline
50 & 0.84 $\pm$ 0.014 & 0.82 $\pm$ 0.005 & 0.74 $\pm$ 0.049 \\ \hline
70 & 0.87 $\pm$ 0.040 & 0.83 $\pm$ 0.032 & 0.76 $\pm$ 0.043 \\ \hline
\end{tabular}}
\caption{The average classification accuracies and standard deviations (\%) for ADNI dataset}
\label{Table_5}
\end{table}
\begin{table*}
\centering
\resizebox{0.60\textwidth}{!}{%
\begin{tabular}{|l|l|l|l|l|l|}
\hline
\multicolumn{1}{|c|}{Algorithm name} & \multicolumn{4}{c|}{UQ-CHI} & \multicolumn{1}{c|}{\multirow{3}{*}{CHI}} \\ \cline{1-5}
\multicolumn{1}{|c|}{\multirow{2}{*}{Label ratio}} & \multicolumn{1}{c|}{\multirow{2}{*}{Training ratio}} & \multicolumn{3}{c|}{Rejection rate} & \multicolumn{1}{c|}{} \\ \cline{3-5}
\multicolumn{1}{|c|}{} & \multicolumn{1}{c|}{} & Low = 20 & Medium = 40 & High = 60 & \multicolumn{1}{c|}{} \\ \hline
\multirow{3}{*}{Low = 10} & 30 & 0.71 & 0.76 & 0.83 & 0.64 \\ \cline{2-6} 
 & 50 & 0.75 & 0.78 & 0.84 & 0.66 \\ \cline{2-6} 
 & 70 & 0.77 & 0.79 & 0.87 & 0.70 \\ \hline
\multirow{3}{*}{Medium = 20} & 30 & 0.67 & 0.71 & 0.72 & 0.58 \\ \cline{2-6} 
 & 50 & 0.70 & 0.72 & 0.75 & 0.62 \\ \cline{2-6} 
 & 70 & 0.71 & 0.75 & 0.76 & 0.63 \\ \hline
\multirow{3}{*}{High = 50} & 30 & 0.66 & 0.70 & 0.71 & 0.55 \\ \cline{2-6} 
 & 50 & 0.69 & 0.71 & 0.73 & 0.58 \\ \cline{2-6} 
 & 70 & 0.71 & 0.72 & 0.74 & 0.62 \\ \hline
\end{tabular}}
\caption{Corresponding testing accuracies for different rejection options for the ADNI dataset}
\label{Table_6}
\end{table*}
\section{Conclusion}\label{concl}
In this paper, we develop the UQ-CHI method to enable uncertainty quantification for continuous patient monitoring. This probabilistic generalization will facilitate a few extensions to the basic CHI model for decision-making purposes. For example, in many degenerative disease conditions such as AD, it is essential to triage patients to determine the priority of resource allocations and patient care. Therefore, the UQ-CHI framework would equip us with an optimal decision considering imperfect and continuous delivery of knowledge. In the future, we would like to extend this method to other diseases that may show different degradation characteristics in the context of degenerative diseases. Another extension of this methodology is to apply on a non-linear index and further explore the feasibility of varying discriminant functions.

 \appendix
 \section{Proof to Lemma \ref{lemma_1}}\label{Appendix_A}
\begin{proof}
By adding a set of dual variables, one for each constraint, the Lagrangian of the optimization problem in \eqref{eq:MED} can be written as:
\begin{equation}\label{lemma_1_lagrangian}
\begin{aligned}
&\mathcal{L}\big(p(\mathbf{w},\lambda)\big) = KL\big(p(\mathbf{w})||p_0(\mathbf{w})\big) - \sum_n \mathbf{\lambda}_n \big(y_n E_{p(\mathbf{w})}[\mathcal{D}\left( X_{n}|\mathbf{w} \right)] - 1\big),
\end{aligned}
\end{equation}
In order to find the solution to Eq. \eqref{eq:MED}, and given the definition of the KL-divergence in \eqref{KL divergence} we require,
\begin{equation}\label{lemma_1_derivative}
\begin{aligned}
&\dfrac {\partial \mathcal{L}}{\partial p\left(\mathbf{w} \right) } = \log p(\mathbf{w}) - \log p_0(\mathbf{w}) - \sum_n \mathbf{\lambda}_ny_n\mathcal{D}\left( X_{n}|\mathbf{w} \right) = 0,
\end{aligned}
\end{equation}
The solution to the MED optimization problem has the following general form:
\begin{equation}\label{lemma_1_solution}
\begin{aligned}
&p(\mathbf{w}^*) = \frac{1}{Z(\mathbf{\lambda})}p_0(\mathbf{w}) \exp \bigg(\sum_n \mathbf{\lambda}_n y_n \mathcal{D}\left( X_{n}|\mathbf{w}\right)\bigg).
\end{aligned}
\end{equation}
Here, $Z(\mathbf{\lambda})$ is the normalization constant defined in \eqref{lemma_1_Z}, then the general exponential form of the solution becomes:
\begin{equation}\label{lemma_1_final proof}
\begin{split}
g(\mathbf{\lambda}) &= \mathcal{L}\big(p(\mathbf{w^*}),\mathbf{\lambda}\big)\\
&= \int \frac{1}{Z(\mathbf{\lambda})}p_0(\mathbf{w})\exp \bigg(\sum_n \mathbf{\lambda}_ny_n \mathcal{D}\left( X_{n}|\mathbf{w} \right)\bigg) \\ 
&\quad \bigg(\sum_n\mathbf{\lambda}_ny_n \mathcal{D}\left( X_{n}|\mathbf{w} \right)-\log p_0(\mathbf{w})-log(\mathbf{Z})\bigg)d\mathbf{w} -\\
&\quad  \sum_n \mathbf{\lambda}_n \Bigg(y_n\int \frac{1}{Z(\mathbf{\lambda})}p_0(\mathbf{w})\exp \bigg(\sum_n \mathbf{\lambda}_ny_n \mathcal{D}\left( X_{n}|\mathbf{w} \right)\bigg)\\
&\quad \mathcal{D}\left( X_{n}|\mathbf{w} \right)d\mathbf{w} - 1 \Bigg)\\
&=\sum_n \lambda_n - \log Z(\mathbf{\lambda}).
\end{split}
\end{equation}
\end{proof}
Hence, the dual of the MED problem can be shown in \eqref{regularZ}. 
\section{Proof to Lemma \ref{lemma_2}}\label{Appendix_B}
\begin{proof}
Let $Z(\mathbf{\lambda})$ be the normalization constant defined in Eq. \eqref{lemma_1_Z}, given the constraints in \eqref{full UQ-CHI} the normalization constant can be reformulated as follows:
\begin{subequations}  \label{eq:MMEDCHIzsol2}
 \begin{align}
& Z(\mathbf{\lambda})=\int p_{0}\left(\mathbf{w,\gamma}\right)\\
&\qquad \quad\exp  \bigg(\sum _{n\in \left\{ 1,\ldots,N\right\} }\mathbf{\lambda}_n[p(y_n) \mathcal{D}(  \mathbf{x}_{n,Tn}\mid \mathbf{w}) - \mathbf{\gamma}_{n}]+   \\
&\quad \quad\sum_{\substack{n\in \{1,\ldots,N\}\\ t\in \left\{ 1,\ldots ,T_{n-1}\right\}} } \mathbf{\lambda}_n\left[\mathcal{M}\left( \mathbf{z}_{n, t}|\mathbf{w} \right) -\mathbf{\gamma}_{n}\right] \bigg) d\mathbf{w} d\mathbf{\mathbf{\gamma}},\\
&\quad \quad=\int P_{0}\left(\mathbf{w}\right)\exp\bigg(\sum_{\substack{n\in \{1,\ldots,N\}} }\mathbf{\lambda}_{n} p_n(y_n)\mathbf{w}^{T}  \mathbf{x}_{n, Tn}+\label{eq:zsol2_b}\\
&\quad\quad\sum_{\substack{n\in \{1,\ldots,N\}\\ t\in \left\{ 1,\ldots ,T_{n-1}\right\}} } \mathbf{\lambda}_n \mathbf{w}^{T}\mathbf{z}_{n,t}\bigg)\label{eq:zsol2_c}\\
&\qquad\quad P_{0}\left( \mathbf{\gamma}\right) \exp \bigg({-\sum_{\substack{n\in \{1,\ldots,N\}\\ } }\lambda_{n} \mathbf{\gamma}_{n}} \bigg)  d\mathbf{w} d\mathbf{\mathbf{\gamma}}\label{eq:zsol2_d},
 \end{align}
\end{subequations}
Given the priors in \eqref{eq:PriorMEDCHI}, each term in Eq. \eqref{eq:MMEDCHIzsol2} can be reformulated as follows: For the term in \eqref{eq:zsol2_b} and \eqref{eq:zsol2_c} we have the followings: 
\begin{equation}\label{eq:pteta}
\begin{aligned}
&Z_\mathbf{w}(\mathbf{\lambda}) = \int P_{0}\left(\mathbf{w}\right)\exp\bigg(\sum_{\substack{n\in \{1,\ldots,N\}} }\mathbf{\lambda}_{n} p_n(y_n)\mathbf{w}^{T}  \mathbf{x}_{n, Tn}+\\
&\qquad \sum_{\substack{n\in \{1,\ldots,N\}\\ t\in \left\{ 1,\ldots ,T_{n-1}\right\}} } \mathbf{\lambda}_n \mathbf{w}^{T}\mathbf{z}_{n,t}\bigg) d\mathbf{w}\\
&\quad \quad=\int ^{\infty }_{-\infty }\frac{1}{\sqrt {2\pi }}\exp\bigg(-\frac{1}{2}\mathbf{w}^T\mathbf{w}\bigg)\exp\bigg(\sum_{\substack{n\in \{1,\ldots,N\}} }\mathbf{\lambda}_{n} p_n(y_n)\mathbf{w}^{T}  \mathbf{x}_{n, Tn} +\\
&\qquad \sum_{\substack{n\in \{1,\ldots,N\}\\ t\in \left\{ 1,\ldots ,T_{n-1}\right\}} } \mathbf{\lambda}_n \mathbf{w}^{T}\mathbf{z}_{n,t}\bigg)d\mathbf{w}\\
&\qquad=\int ^{\infty }_{-\infty }\frac{1}{\sqrt {2\pi }}\exp\Bigg(-\frac{1}{2}\bigg(\mathbf{w}^T\mathbf{w} - 2\mathbf{w}^T\big(\sum_{\substack{n\in \{1,\ldots,N\}} }\mathbf{\lambda}_{n} p_n(y_n)  \mathbf{x}_{n, Tn}+ \\
&\qquad \sum_{\substack{n\in \{1,\ldots,N\}\\ t\in \left\{ 1,\ldots ,T_{n-1}\right\}} } \mathbf{\lambda}_n\mathbf{z}_{n,t}\big)\bigg)\Bigg) d\mathbf{w}\\
&\qquad=\exp \Bigg (\frac{1}{2}\bigg(\sum_{\substack{n\in \{1,\ldots,N\}} }\mathbf{\lambda}_{n} p_n(y_n)  \mathbf{x}_{n,Tn} +\sum_{\substack{n\in \{1,\ldots,N\}\\ t\in \left\{ 1,\ldots ,T_{n-1}\right\}} } \mathbf{\lambda}_n \mathbf{z}_{n,t}  \bigg)^T\\
&\qquad\quad\bigg(\sum_{\substack{n\in \{1,\ldots,N\}} }\mathbf{\lambda}_{n} p_n(y_n)  \mathbf{x}_{n,Tn} +\sum_{\substack{n\in \{1,\ldots,N\}\\ t\in \left\{ 1,\ldots ,T_{n-1}\right\}} } \mathbf{\lambda}_n \mathbf{z}_{n,t}\bigg)\Bigg)\\
&\qquad\quad\int^{\infty }_{-\infty }\exp\Bigg(-\frac{1}{2}\bigg(\mathbf{w}-\big(\sum_{\substack{n\in \{1,\ldots,N\}} }\mathbf{\lambda}_{n} p_n(y_n)  \mathbf{x}_{n,Tn} +\\
&\qquad\sum_{\substack{n\in \{1,\ldots,N\}\\ t\in \left\{ 1,\ldots ,T_{n-1}\right\}} } \mathbf{\lambda}_n \mathbf{z}_{n,t}\big)\bigg)^T\\
&\qquad\quad\bigg(\mathbf{w}-\big(\sum_{\substack{n\in \{1,\ldots,N\}} }\mathbf{\lambda}_{n} p_n(y_n)  \mathbf{x}_{n,Tn} +\sum_{\substack{n\in \{1,\ldots,N\}\\ t\in \left\{ 1,\ldots ,T_{n-1}\right\}} } \mathbf{\lambda}_n \mathbf{z}_{n,t}\big)\bigg)\Bigg)\\
&\qquad=\exp \Bigg (\frac{1}{2}\bigg(\sum_{\substack{n\in \{1,\ldots,N\}} }\mathbf{\lambda}_{n} p_n(y_n)  \mathbf{x}_{n,Tn} +\sum_{\substack{n\in \{1,\ldots,N\}\\ t\in \left\{ 1,\ldots ,T_{n-1}\right\}} } \mathbf{\lambda}_n \mathbf{z}_{n,t}  \bigg)^T\\
&\qquad\quad\bigg(\sum_{\substack{n\in \{1,\ldots,N\}} }\mathbf{\lambda}_{n} p_n(y_n)  \mathbf{x}_{n,Tn} +\sum_{\substack{n\in \{1,\ldots,N\}\\ t\in \left\{ 1,\ldots ,T_{n-1}\right\}} } \mathbf{\lambda}_n \mathbf{z}_{n,t}\bigg)\Bigg),
\end{aligned}
\end{equation}

And for the last term \eqref{eq:zsol2_d} we have the following: 
\begin{equation}\label{eq:pgamma}
\begin{aligned}
Z_\mathbf{\gamma}(\mathbf{\lambda}) &= P_{0}\left( \mathbf{\gamma}\right) \exp \big({-\sum_{\substack{n\in \{1,\ldots,N\}\\ } }\lambda_{n} \mathbf{\gamma}_{n}} \big)  d\mathbf{\mathbf{\gamma}}\\
&=\prod ^{n}_{1}\int  P_{0}\left( \mathbf{\gamma}_{n}\right) \exp \big({-\lambda_{n} \mathbf{\gamma}_n} \big) d\mathbf{\gamma}_n\\
&=\int ^{1}_{-\infty } c \exp\big(-c+c\mathbf{\gamma}_n\big)\exp \big(-\lambda_{n}\mathbf{\gamma}_n \big)d\mathbf{\gamma}_n\\
&=\frac{c}{c-\mathbf{\lambda}_n}\exp\big(-\lambda_n\big)\\
&= \frac{1}{1-\mathbf{\lambda}_t/c}\exp\big(-\mathbf{\lambda}_n\big),
\end{aligned}
\end{equation}

Substituting the results from Eq. \eqref{eq:pteta} and \eqref{eq:pgamma}  in \eqref{eq:MMEDCHIzsol2}, results in Eq. \eqref{finalZZ}.
\end{proof}
\section{Proof to Lemma \ref{lemma_3}}\label{Appendix_C}
\begin{proof}
Given the marginal distribution $p\left(\mathbf{w}\right)$ in \eqref{distributionestimate} and the convex combination of discriminant functions defined as $\int p(\mathbf{w})\mathcal{D} \left(  \mathbf{x}|\mathbf{w}\right)d\mathbf{w}$ we have the following:
\begin{equation} \label{appenc}
\begin{split}
\int p(\mathbf{w})\mathcal{D}\left(   \mathbf{x}|\mathbf{w} \right)d\mathbf{w} &=\int p(\mathbf{w})(\mathbf{w}^T  \mathbf{x}_{new} )d\mathbf{w}\\
&=\frac{1}{Z_\mathbf{w}(\mathbf{\lambda})}\mathbf{w}^T  \mathbf{x}_{new}
p_{0}\left( \mathbf{w}\right)\\
&\quad\exp \bigg(\sum_{\substack{n\in \{1,\ldots,N\}}}\mathbf{\lambda}_n p_n(y_n)\mathbf{w}^T  \mathbf{x}_{n,Tn}+ \\
&\sum _{\substack{n\in \{1,\ldots,N\}\\ t\in \left\{ 1,\ldots ,T_{n-1}\right\}}}\mathbf{\lambda}_{n} \mathbf{w}^T\mathbf{z}_{n, t}\bigg) d\mathbf{w},
\end{split}
\end{equation}
Where, given the prior distributions, \eqref{appenc} can be written as:
\begin{equation}
\begin{aligned}
&=\int ^{\infty }_{-\infty }\frac{\mathbf{w}^T  \mathbf{x}_{new}}{Z_{\mathbf{w}}(\mathbf{\lambda})\sqrt {2\pi }}\exp \Bigg(-\frac{1}{2}\mathbf{w}^T\mathbf{w}+ 
\mathbf{w}^T \bigg(\sum_{\substack{n\in \{1,\ldots,N\}}}\mathbf{\lambda}_n p_n(y_n)  \mathbf{x}_{n,Tn} + \\
&\sum _{\substack{n\in \{1,\ldots,N\}\\ t\in \left\{ 1,\ldots ,T_{n-1}\right\}}}\mathbf{\lambda}_{n} \mathbf{z}_{n, t}\bigg)\Bigg)d\mathbf{w}\\
&=\int ^{\infty }_{-\infty }\frac{\mathbf{w}^T  \mathbf{x}_{new}}{\sqrt {2\pi }}\exp\Bigg(\quad\frac{-1}{2}\bigg(\mathbf{w} - \big(\sum_{\substack{n\in \{1,\ldots,N\}}}\mathbf{\lambda}_n p_n(y_n)  \mathbf{x}_{n,Tn} +\\
&\sum _{\substack{n\in \{1,\ldots,N\}\\ t\in \left\{ 1,\ldots ,T_{n-1}\right\}}}\mathbf{\lambda}_{n} \mathbf{z}_{n, t}\big)\bigg)^T\bigg(\mathbf{w} - \big(\sum_{\substack{n\in \{1,\ldots,N\}}}\mathbf{\lambda}_n p_n(y_n)  \mathbf{x}_{n,Tn} + \\
&\sum _{\substack{n\in \{1,\ldots,N\}\\ t\in \left\{ 1,\ldots ,T_{n-1}\right\}}}\mathbf{\lambda}_{n} \mathbf{z}_{n, t}\big)\bigg)\Bigg)d\mathbf{w}\\
&=\Bigg(\sum_{\substack{n\in \{1,\ldots,N\}}}\mathbf{\lambda}_n p_n(y_n)  \mathbf{x}_{n,Tn} + \sum _{\substack{n\in \{1,\ldots,N\}\\ t\in \left\{ 1,\ldots ,T_{n-1}\right\}}}\mathbf{\lambda}_{n} \mathbf{z}_{n, t}\Bigg)^T  \mathbf{x}_{new}.
\end{aligned}
\end{equation}
\end{proof}

\section*{Conflict of interest}
The authors declare that they have no competing interests.

\section*{Acknowledgements}
The authors acknowledge funding support from the National Science Foundation under Grants CMMI-1536398 and CCF-1715027.

\bibliographystyle{elsarticle-num} 
\bibliography{Reference}
\end{document}